# Crystal Hybridized Pyro-Piezoelectric Ferrofluidic Harvester

**Abdullahi Isse**
University of Minnesota
116 Church St SE, Minneapolis, MN, 55455, United States

***Abstract*** - This paper investigates the usage of a hybridized Pyro-Piezo Ferrofluidic Harvester to capture useful energy from ambient vibration and temperature fluctuation. Consisting of a ferrofluidic core, the proposed device utilizes pyro and piezoelectric materials to charge an integrated ultracapacitor for 'trickle' energy harvesting and usage. In particular, a wearable or implantable hybridized crystal quartz piezoelectric and PZT pyroelectric based ferrofluidic harvester is characterized under human ambient energy boundaries. The frequency associated with typical bodily movement such as walking and strenuous bodily movement such as running was determined via accelerometer and analyzed via fast fourier transform, with determined mean frequencies of 1.7 Hz for typical bodily movement and 4.3 Hz for strenuous movement. Under an optimal continuous harvestation period of ~24 hours, the CQF harvester was characterized with a energy harvestation of 7.89 MJ.

***Keywords***: Ferroelectricity, Pyroelectricity, Piezoelectricity, Ferrofluids, Energy, Harvesting

## 1. Introduction

Ferrofluids are magnetizable colloid solutions consisting of conglomeration preventing surfactant coated iron-oxide nanoparticles immersed in an organic solvent. In particular, these nanoparticles range from size of 5 nm to 20 nm, at such scales the nanoparticles are characterized by an apparent magnetic mono-domain in which subsequently results a state of superparamagnetism[1]-[2]. In which allows the nanoparticles or ferrofluid to be magnetizable upon an applied external magnetic field with a greater associated magnetic suspectiability due to the nanoparticle's unidirectional internal magnetization characterized by the magnetic moment μ that is proportional to the material-dependent saturation magnetization $M_s$ and particle volume V

$$\mu = M_s V \qquad (1)$$

Ferroelectricity is the effect describing the phenomena of crystal materials with an observed non-zero spontaneous electric polarization that is additionally reversible under an applied external electric field, this effect is conserved across materials characterized by non-centrosymmetric lattice systems. In particular, the crystal materials internal polarization is unit-cell dependent-as a result any change in a lattice structure induces a change in dipole strength and thus a change in internal polarization that subsequently results in an induced surface charge[3]-[4]. Depending on a particular crystal lattice system, the pyroelectric and piezoelectric effect is additionally exhibited. Pyroelectricity refers to the phenomenon of a material developing a voltage due to a change in temperature. In particular, temperature fluctuations induces lattice system reorientation within the crystal in which subsequently results in a change of internal polarization. Under conditions of uniform temperature change across the material,the piezoelectric effect is characterized by the material-dependent pyroelectric coefficient Ÿ. In certain non-centrosymmetric materials an applied external mechanical stress will induce a change in crystal polarization due to the change in the dipole coupled lattice structure, similar to the pyroelectric effect-this results in a surface charge across the material. The converse additionally occurs, an applied voltage across the material results in a crystal deformation proportional to the applied electric field[5]-[6]. Similarly, the piezoelectric effects behavior is characterized by the material-dependent piezoelectric coefficient ϱ.

## 2. Theoretical

The ferrofluidic motion in the cylindrical core can be characterized as a variable-mass system, represented by

$$F_{ext} + v_{rel}\frac{dm}{dt} = mx'' \qquad (2)$$

Whereas m is the mass under the zone of equilibrium and $v_{rel}$ is the relative velocity of the particles outside the zone of equilibrium. Here I define the zone of equilibrium as the observed corrugated geometry of the ferrofluid under conditions of an applied constant external magnetic field. In particular, normal-field instability is responsible for this apparent corrugation. As a consequence of the concentrated external magnetic field, the stabilization of magnetic potential competing with interfacial and gravitational potential results in the observed peak-valley surface of the ferrofluid [7]-[8]. In addition, the ferrofluidic motion is also modeled as a spring-mass system with an associated damping constant as a function of drag and spring constant as a function of magnetic potential.

$$mx'' + cx' + kx = F_{ext} \qquad (3)$$
$$c = \zeta c_c \qquad (4)$$
$$k = -\mu B \qquad (5)$$

Equation (2) can be related to equation (3), whereas a constant α term is utilized to consider for non-linear behaviors within the range of 0.7 to 5 Hz.

$$\alpha(mx'' + cx' + kx) = mx'' - v_{rel}\frac{dm}{dt} \qquad (6)$$

By considering only surface movement across x-axis, the changing ferrofluidic motion in the core will induce an electromotive force along the adjacent coiling via Faraday's law of induction

$$\varepsilon = -N\frac{d}{dt}\int_{dz}\int_{dx} B(x)\cos(\omega t)dxdz \qquad (7)$$

As described by the pyroelectric effect, a material's charge displacement is proportional to the area and temperature fluctuation across the material, represented by the following relation

$$Q(t) = \Upsilon A \Delta T(t) \tag{8}$$

With an output current expressed by

$$I(t) = \Upsilon A \frac{dT}{dt} \tag{9}$$

The final component of the general PPF harvester to characterize is piezoelectricity, in which a materials charge displacement is proportional to the applied external force across the material, represented by

$$Q(t) = \varrho \Delta F(t) \tag{10}$$

With an associated output current expressed as

$$I(t) = \varrho \frac{dF(t)}{dt} \tag{11}$$

As a central theme of simplicity and functionality, namely a device with little to no moving parts; I construct a circuit directly corresponding to the physical prototype.

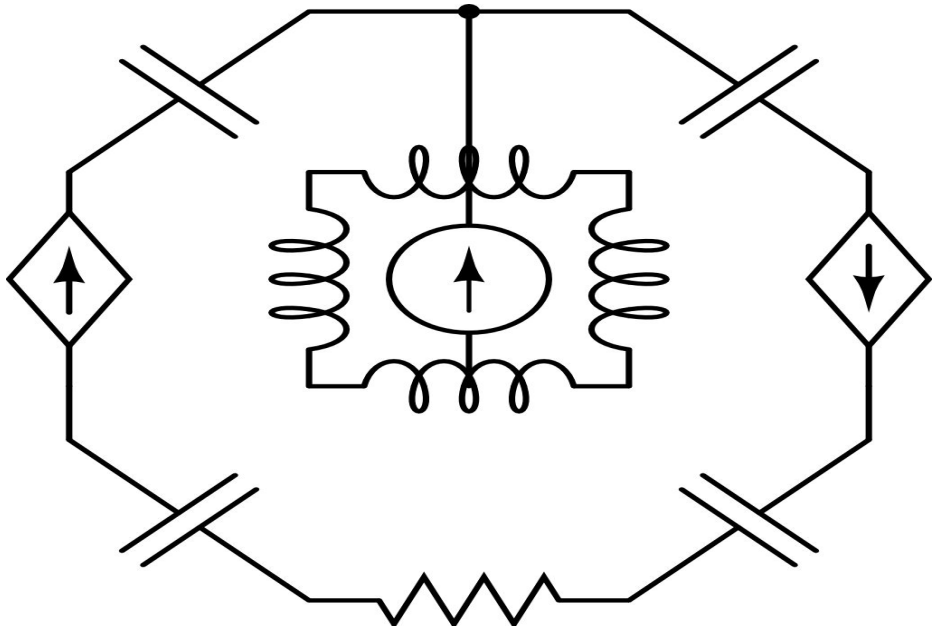

Figure 1: General Pyro-Piezoelectric Ferrofluidic Harvester (PPFH) Circuit Schematic

The ferrofluidic core and pyroelectric componetent(s) are modeled as an independent current source , whereas the piezoelectric component(s) are modeled as a capacitor in the schematic for engineering parameter characterization but will be treated as an additional current source as expressed by

$$I(t) = \varrho \omega F_0 \cos(\omega t) \tag{12}$$

The total effective current across the load resistor is obtained by summing the individual current contributions of each component(s) for total current $I_T$ with an associated system power output

$$P = (I_f + I_\Upsilon + I_\varrho)^2 R \tag{13}$$

Whereas $I_f$ is the current produced by cylindrical ferrofluidic core, $I_\ddot{\Upsilon}$ the current produced by the piezoelectric component(s) and $I_\varrho$ the current produced by the pyroelectric component(s)

## 3. Results & Discussion

Towards the development of an implantable or wearable general PPF harvester, the ambient energy of the human body must be characterized. In the case of the PPF harvester, this would be the frequency associated with a wide range of bodily movement and the temperature fluctuations of the body throughout the day. Utilizing standard accelerometers, the acceleration data associated with walking was measured for a total of 5 minutes with a sample rate of 10 s as well as the acceleration associated with running under similar conditions.

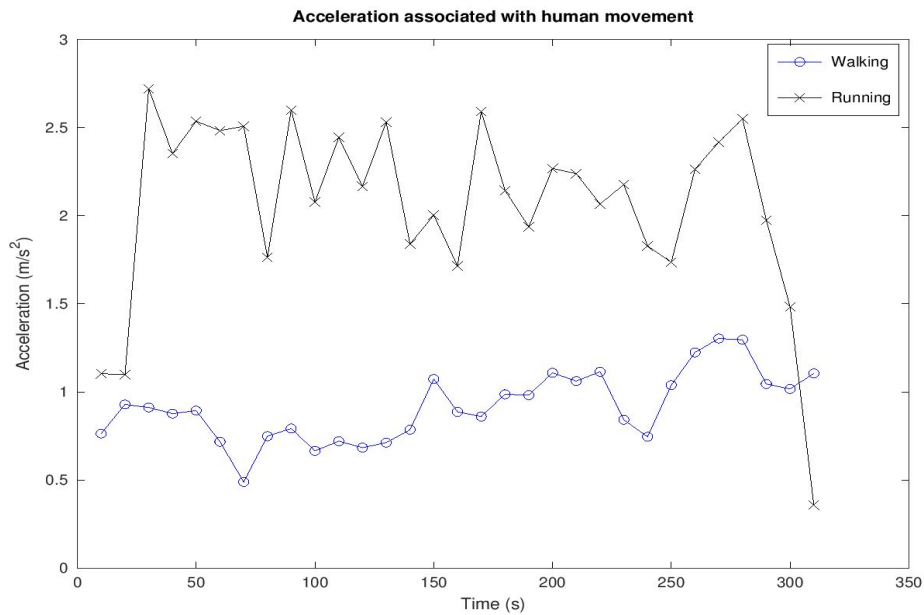

Figure 2: Acceleration associated with a range of human motion, whereas 'walking' includes climbing stairs, sitting and whereas 'running' includes jumping.

The frequency associated with human movement was determined across two modes: a passive mode representing typical movements of daily tasks, primarily walking and an active mode representing movements associated with high physical exertion such as running or playing sport. The acceleration data across the two modes as seen in figure (3) was analyzed in the frequency domain via the fast fourier transform [9] with a sample rate of 10 s utilizing Octave software [10].

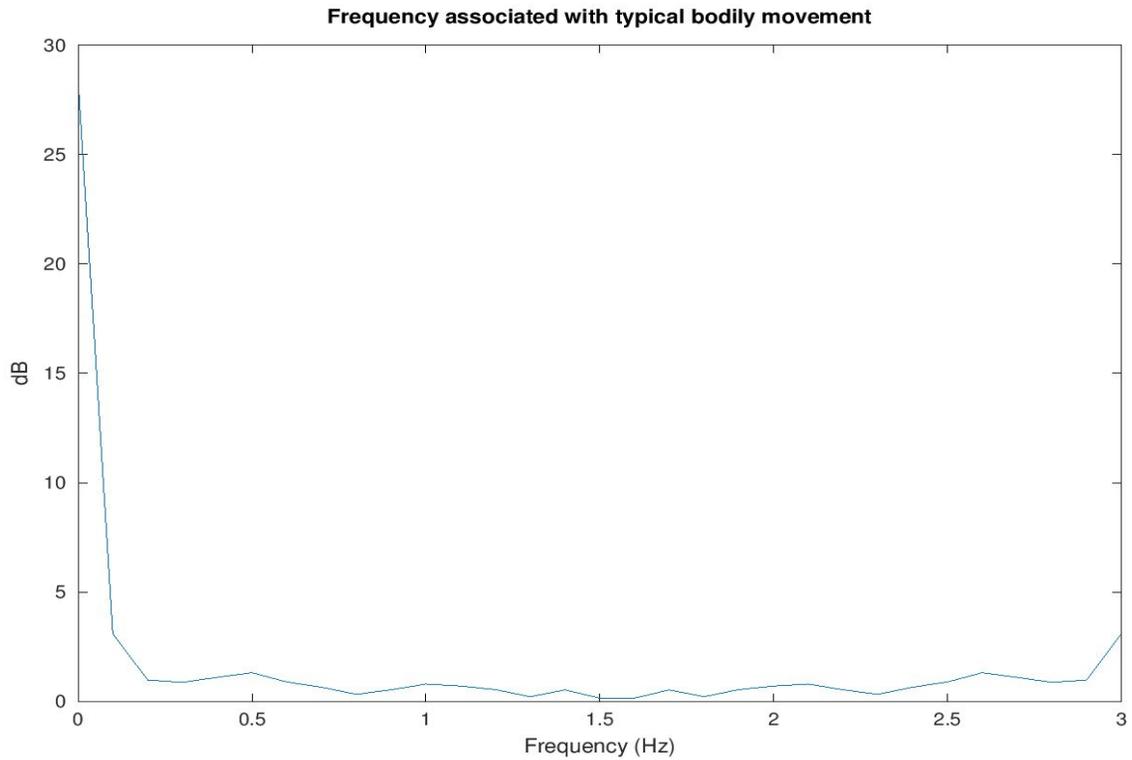

Figure 3: Frequency associated with typical bodily movement such as walking.

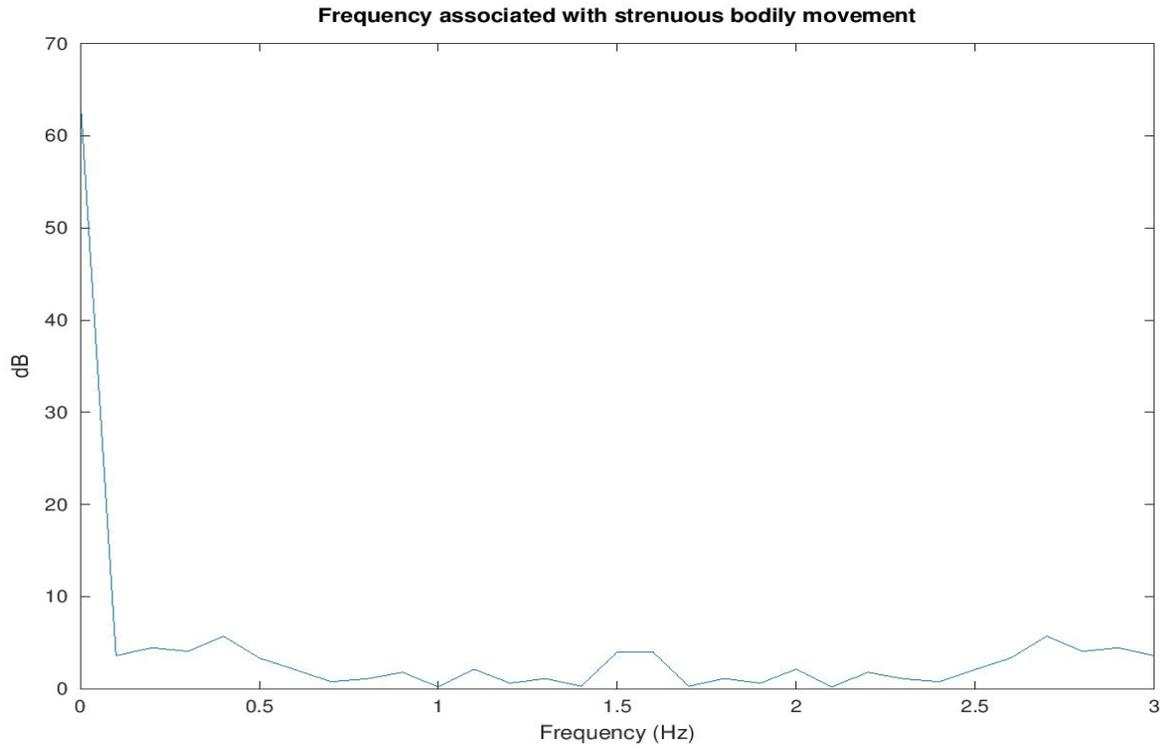

Figure 4: Frequency associated with strenuous bodily movement such as running

Although body temperature varies throughout the day due to the circadian rhythm or varies across individuals, to sustain proper health the body maintains a relatively constant internal body temperature at 37 C [11,12,13]. As such, the pyroelectric componetent of the CQF harvester must utilize another method of developing a temperature change. One method to induce a temperature change is by capitalizing on the heat produced by conductive wiring due to the ferrofluidic induced current flow as expressed by Joule's First law

$$q = I^2 R t \qquad (14)$$

Equation (14) can be related to the specific heat capacity formula to create an expression for the temperature change across the conductive wire

$$\Delta T = \frac{I_f^2 R t}{mc} \qquad (15)$$

By plugging in equation (14) as a function of temperature change into equation (8), the pyroelectric component is characterized with a current output of

XXX-6

$$I(t) = \Upsilon A \frac{I_f^2 R}{mc} \tag{16}$$

Table 1: Select properties of pyro and piezoelectric crystals

| Crystal | Formula | $\Upsilon$ (C/cm²K) | $\varrho$ (cm/V) | Temperature range (K) | Density (g/cm³) |
|---|---|---|---|---|---|
| Silicon Dioxide (Quartz) | $SiO_2$ | N/A | $2.25 \times 10^{-10}$ | N/A | 2.65 |
| Deuterated Triglycine Sulfate (DTGS) | $C_6D_{17}N_3O_{10}S$ | $3 \times 10^{-8}$ | N/A | 243-334 | 1.69 |
| Lithium Tantalate (LT) | $LiTaO_3$ | $1.9 \times 10^{-8}$ | $3 \times 10^{-10}$ | 273-891 | 7.46 |
| Lead Zirconate Titanate (PZT) | $Pb(Zr_{0.52}Ti_{0.48})O_3$ | $5.5 \times 10^{-8}$ | $2.34 \times 10^{-8}$ | 298-523 | 7.70 |
| Barium Titanate (BT) | $BaTiO_3$ | $3.3 \times 10^{-8}$ | $1.9 \times 10^{-8}$ | 293-320 | 5.70 |
| Polyvinyl Chloride (PVC) | $(C_2H_3Cl)_n$ | $0.4 \times 10^{-9}$ | $0.7 \times 10^{-10}$ | N/A | 1.40 |
| Polyvinyl Fluoride (PVF) | $(C_2H_3F)_n$ | $1.8 \times 10^{-9}$ | $1 \times 10^{-10}$ | N/A | 1.45 |
| Polyvinylidene Fluoride (PVDF) | $(C_2H_2F_2)_n$ | $4 \times 10^{-9}$ | $4 \times 10^{-10}$ | N/A | 1.78 |

Although equations (2-6) provide a novel and complete pathway from first principles, it requires further investigation outside the scope of this paper to be validated experimentally. As such, I will utilize the ferrofluidic power density data from previous investigations as a case example [14], which determined an associated power density of 8 μW/g within the range of 2-2.2 Hz-which is validated within the characterized frequency range. Utilizing the circuit schematic from figure (1) as a model, the power output of the crystal qaurtz ferrofluidic harvester (CQFH) can be determined by summing each individual current source across the resistor load, whereas an integrated ultracapacitor will be modeled as the load resistor with an esr of 0.09 mΩ. Under a frequency range of 2.0-2.4 Hz, the cylindrical ferrofluidic core under dimensions of a 76.2 mm radius with 127 mm height wrapped with 1000 turns of coil contributes 0.0114 amps of current, the quartz piezoelectric component 2.4 nm amps and under conditions of 1 hour ferrofluidic core operation the PZT pyroelectric component 3.03 nm amps for a total current of ~.0114 amps across the load resistor. Based on initial data calculations characterized under human motion and heat fluctuation based on ferrofluidic core coilings, the only useful current source is from the ferrofluidic core at the range of ~1 hour operational time. To overcome this issue, I propose the use of a ultracapacitor integrated into the CQFH. In particular, as the CQFH continuously produces power assuming bodily movement or temperature fluctuations occur, this power can be trickled into a ultracapacitor with an esr of 0.09 mΩ for a maximum energy harvestatation of 7.89 MJ under conditions of continuous 24 hour optimal operation. The practical maximum energy harvestation is dependent on CQFH size constraints, ultracapacitor and pyroelectric thermal cycling parameters [15].